# A Short Report on a Probability-Based Interpretation of Quantum Mechanics

Paolo Rocchi
IBM, via Luigi Stipa 150, Roma, Italy;
LUISS University, via Romania 32, Roma, Italy;  procchi@luiss.it

**Abstract**: This paper calls attention to the current state of the probability (*P*) domain which presents weak points at the mathematical level and more significant flaws at the application level. Popper notices how fundamental issues raised in quantum mechanics (QM) directly derive from unresolved probabilistic questions. Endless philosophical debates create more problems than solutions, so the author of this research suggests going directly to the root of the issues and searching for the probability theory which formalizes the multifold nature of *P*.
This paper offers a brief overview of the *structural theory of probability*, recently published in a book, and applies it to QM in order to show its completeness. The whole *probability-based interpretation of QM* goes beyond the limits of a paper and these pages condense a few aspects of this theoretical scheme. The double slit experiment is used to corroborate the theorems presented here.

## 1 Introduction

Most quantum scientists trust in the correctness of the probability calculus and deem the contrasts amongst the probability schools are no more than philosophical quarrels. In reality, things are not exactly in these terms. Let us go through some details.

■ Kolmogorov grounds his construction on the nonnegativity, normality and additivity axioms. Most mathematicians share this theoretical base which however does not characterize probability in exhaustive terms [1].

■ We read in [2]: "If $P(A) > 0$ then the quotient:

$$P_A(B) = \frac{P(AB)}{P(A)}. \qquad (1)$$

is defined to be the *conditional probability* of the event *B* under the condition *A*." This formulation lacks a conceptual justification; as second, it cannot be applied everywhere; as third, the relationship with the notion of *independence* remains inexplicit and some mathematician concludes that in substance (1) is a hidden axiom [3].

■ Eventually, probability theorists are aware of the *single case problem* dealing with the probability of the individual case, which diverges from material evidence [4][5]. However, they have confronted that problem from the philosophical viewpoint rather than with analytical-mathematical methods.

Issues are even more serious as regard the use of *P* in applications. Probability is a parameter employed in countless fields which have inspired different formulations and interpretations. Laplace puts forward the first definition, sometimes called 'classical', referring to equally likely events. He states that the probability of the event *A* is the fraction *f* of the total number of possibilities *T* in which *A* occurs:

$$P(A) = \frac{f}{T}. \qquad (2)$$

Gamblers, actuaries and scientists have long understood that relative frequencies bear an intimate relationship to probabilities. The *frequency interpretation* assumes *P* as the limiting relative frequency in suitable infinite sequences of trials [6]. The *propensity interpretation*, originated with Popper [7], regards probability as an objective feature; namely, probability is thought of as a physical disposition, or tendency of a system to produce given outcomes [8]. *Subjectivist* and *Bayesians* define the probability of *A* as the degree of an agent's credence or expectation, about *A* based on prior information [9]. The agent is an ideal rational individual respecting precise rules. Obviously, the degree of belief raises doubts when it refers to objective and testable phenomena in physics especially when it should explain quantum properties [10]. Logicians relate probability to the propositions expressing the premise and the conclusion of reasoning. The *logical interpretation* developed by Carnap [11] takes probability as an extension of *inductive logic*, namely *P* assesses the degree of confirmation of certain evidence. For Keynes probability expresses the degree of *logical implication* between a hypothesis and the rational conclusion, probability is an evidential relation between two statements [12].

The irreconcilable models of probability lead to irreconcilable statistical methods. In fact, frequency interpretation is the position that lies in the background of classical statistics; and subjectivist interpretation underpins the Bayesian statistics. The two statistical schools sometimes suggest similar techniques however they provide answers that have incompatible contents [13].

Concluding, *every probability theory is incomplete* since (**α**) it focuses on a partial aspect of *P*, (**β**) constructs assume insufficient axioms, (**γ**) some formulas are not proven or justified, (**δ**) the single case problem has not been resolved in mathematical terms. Actually, no formulation is currently accepted as definitive by the scientific community.

## 2 Fallout in QM

The weak points (**α**), (**β**), (**γ**) and (**δ**) inevitably affect the studies of quantum mechanics intrinsically rooted in probability. Let us confine ourselves to three points.

● The study of a large sample size of quanta seems to be more viable, while the individual quantum mirroring the single case problem opposes the greatest difficulties:

> "The attempt to conceive the quantum-theoretical description as the complete description of the individual systems leads to unnatural theoretical interpretations, which become immediately unnecessary if one accepts the interpretation that the description refers to ensembles of systems and not to individual systems" [14].

● Max Born referred his theory of collisions to a set of quanta which has statistical properties [15], but he did not develop a more detailed description of that set. The followers of the *ensemble interpretation* underscore that the experimental control of probability requires a set of multiple data, that is a vast multitude of quantum systems subjected to similar mechanical conditions [16]. Thus, the wavefunction should not be applied to an individual system, but to an ensemble of particles. This approach, which conforms to the frequentist perspective, follows the diametrically opposite direction of *Quantum Bayesianism* (Qbism) that draws directly from the Bayesian school [17]. Qbism holds that the main aspects of the quantum formalism are subjective in nature; in particular, a quantum state is not an element of reality, but it represents the degrees of belief an agent has about the possible outcomes of measurements [18]. The followers of Qbism deny the criticism about unrealism because the participation of the observer to the measurement process could be associated with a kind of realism they call "participatory realism".

● An isolated quantum system evolves in time in a deterministic way according to the Schrödinger equation and the rule postulated by Born [16] associates the squared wavefunction to probability:

$$P(x, y, z, t_0) = |\psi(x, y, z, t_0)|^2. \qquad (3)$$

The wavefunction and other open probability arguments give rise to dozens of quantum interpretations. For example. for the Copenhagen school, the gap between determinism and indeterminism arises from the "irreducible indeterminacy" of quantum physics; for Einstein, it is logical to assume that finer and subtler processes, which are therefore hidden, interfere [19]. Probability theories are incomplete and do not provide great help especially for the following issues [20]:
   **i.** The *wave/particle nature of quanta*,

ii.   The *collapse of the wave* and
   iii.  The *measurement process,*

Defects (**α**), (**β**), (**γ**) and (**δ**) prevent clarifying the aleatory phenomena in ordinary environments, they do not clarify the phenomena in the quantum environment either. Popper concludes that only a comprehensive construction will provide solid answers. I shared his lesson and have searched for a unified framework about probability.

After some preliminary reports [21][22][23], the recently published book [24] gives the full account of this inquiry; it presents all the definitions, proves fifteen theorems and discusses fourteen experiments. Such a complex proposal goes beyond the limit of this paper which instead includes the ensuing parts:

   **(A)** It summarizes the novel probability framework and tackles the single case problem (point **δ**).
   **(B)** It imports the theoretical results in QM and provides original answers to issues **i, ii** and **iii**.

## 3 Research strategy

The new proposed theory is not axiomatic (point **β**), but is grounded on the accurate analysis of the object measured by *P*.

1)   In science a parameter takes on different meanings in correspondence of the gauged entities. E.g. the price *p*(*G*) is a *cost* if the item *G* is purchased; it is a *gain* if *G* is sold, namely *p*(*G*) takes on opposed attributes depending on *G*. Each probability school assesses a different object, for instance:
   - Laplace computes equally likely events.
   - Von Mises studies phenomena formed by many repeated events of the same type.
   - The subjectivists and the Bayesians refer *P* to individual credence.
   - Keynes means to quantify rational reasoning.
   - Carnap draws conclusions from evidence and qualifies inductive reasoning.

Therefore, probability *P*(*E*) takes on a set of meanings in consequence of the different arguments *E*, and to untangle the riddles raised by the multifold nature of probability we delve into *E*.

2)   Every probability theoretical framework proves to be effective within a specific area. This entails that the comprehensive theory should calculate all the objects listed in 1).

3)   In consequence of remark 2) the multifold argument *E* should be formulated with great precision. This method does not match with the majority of coeval authors who share the simplified model of *E*. They assume that *the event E is a subset of results* belonging to the event space:

$$E = \{\omega_1, \omega_2, \ldots\} \qquad (4)$$

This solution raises the following doubts:
- Definition (4) hints that *E* and ω should share the same nature, conversely recent studies [25][26] and universal experience show how they are very close but distinct. The event is the process that begins with certain initial conditions and ends with the emission of the result. The former is the overall happening, and the result is a part of it.
- Subjectivists, Bayesians and logicians prefer to describe the event by means of sentences which badly fit with the set model (4).
- The *initial conditions* are essential to identify certain, uncertain and impossible events; yet (4) gives this notion for granted.

Concluding, the comprehensive conceptualization of probability should not give up the faithful description of the event, and for this purpose, we adopt the following *structure* including *the elements* α, ω *and the relation* ρ connecting them [27]. In detail, ρ formalizes the process which brings about to the outcome ω from the initial conditions α:

$$\mathbf{E} = (\alpha, \rho, \omega). \qquad (5)$$

Where the result ω can be expressed by a subset or even a proposition.

The structure (5) describes anything that happens: a material phenomenon and also a reasoning, a deduction, a credence etc. which are mental events. This theory embraces the diverse interpretations of *P* by means of (5). For example, raining is a material occurrence caused by various meteorological factors and has this structure:

$$\mathbf{E_p} = (\textit{meteorological factors, falling, rain}). \tag{6}$$

Tom examines the sky and concludes it will rain; this uncertain reasoning can be expressed this way:

$$\mathbf{E_m} = (\textit{meteorological information, credence, 'it will rain'}). \tag{7}$$

The term '*rain*' indicates the material outcome of the physical event $\mathbf{E_p}$, and the sentence *'it will rain'* is the logical conclusion of the mental $\mathbf{E_m}$. Book [24] analyzes Bayesian, logical and other intellectual events in detail.

The accurate model (5) defines any event which happens, will happen or potentially happens; and probability quantifies this capability to occur no matter $\mathbf{E}$ is physical or intellectual:

$$0 \leq P(\mathbf{E}) \leq 1. \tag{8}$$

E.g. the objective probability of (6) qualifies a fact occurring in the world; the Bayesian probability of (7) assesses a subjective belief developed in the mind of an individual.

Ordinarily, the output ω marks the end of the event, ω determines whether $\mathbf{E}$ occurs or does not occur; hence the event and its outcome share the same probability value:

$$P(\omega) = P(\mathbf{E}). \tag{9}$$

From now onward we delve into physical events whose relative frequency *F* is the experimental observable corresponding to *P* calculated in abstract. Model (5) allows us to locate any material happening in the time scale; $\mathbf{E}$ starts at $t = 0$ and finishes at $t_\omega$ that is the delivery time of ω. In case of repeated trials, $t_\omega$ is the delivery time of the last trial. We call the *time intervals* $T_1$ with $0 < t < t_\omega$, and $T_2$ with $t \geq t_\omega$. We define the *determinate status* $\omega^{(D)}$ of the result whose probability or frequency are integers:

$$P(\omega) = 0; \quad F(\omega) = 0,$$
$$P(\omega) = 1; \quad F(\omega) = 1. \tag{10}$$

And the *indeterminate status* $\omega^{(I)}$ when one of the followings is true:

$$0 < P(\omega) < 1; \quad 0 < F(\omega) < 1. \tag{11}$$

We distinguish the event $\mathbf{E_1}$ that occurs only once, from *the long-term event* $\mathbf{E_\infty}$ that repeats indefinitely $\mathbf{E_1}$:

$$\mathbf{E_\infty}^{(I)} = (\mathbf{E_{11}}^{(I)}, \mathbf{E_{12}}^{(I)}, \mathbf{E_{13}}^{(I)}, \ldots). \tag{12}$$

The unlimited series of single outcomes that are independent and identically distributed, make the outcome of $\mathbf{E_\infty}$ also called *collective*:

$$\omega_\infty^{(I)} = (\omega_{11}^{(I)}, \omega_{12}^{(I)}, \omega_{13}^{(I)}, \ldots). \tag{13}$$

## 4 Theorems

We briefly recall five theorems proved in [24].

**4.1** The *theorem of a single number* (TSN) demonstrates that the relative frequency does not match with probability:

$$F(\mathbf{E_1}) \neq P(\mathbf{E_1}). \tag{14}$$

The probability of a single event never and ever can be directly controlled using experiments and (14) begins to formalize the single case problem which instead thinkers have treated through so many philosophical commentaries.

**4.2** The frequentists judge non-sensical $P(\mathbf{E_1})$ and focus on $P(\mathbf{E_\infty})$. In fact, the *theorem of large numbers* (TLN) proves that the relative frequency approaches the probability when the number of trials tends to infinity:

$$F(\mathbf{E_n}) \xrightarrow{a.s.} P(\mathbf{E_\infty}), \qquad \text{as } n \to \infty. \tag{15}$$

This result ensures that $P(\mathbf{E_\infty})$ is a physical and testable quantity at least in principle. The reader should keep (15), that is the law of large numbers in the Borel form, apart from the *weak and strong laws of large numbers* (LLN) that give the account of *the statistical convergence of empirical data toward the expected value.* TLN and LLN are close but have different contents and pursue different scopes.

**4.3** Even subjectivists and Bayesians are aware that the probability of a material occurrence is out of experimental control, so they assume that $P$ assesses not a physical event but an individual's credence about that event. Each individual reasons on the basis of prior information, knowledge or experience α and behaves according to consistent rules to arrive at making the decision, prevision or conclusion ω. The subjectivist scheme allows for probability problems of any kind, even hypothetical, while the *exchangeability theorem* deals with a series of trials. Bayesian statistics provide methods for determining and even updating the degree of personal belief measured by $P(\mathbf{E_1})$.

Result (14) assumes the number of trials $n = 1$, (15) hypothesizes $n \to \infty$, the conditions of TSN and TLN do not overlap and thus the frequentist and subjective model cohabit without conflict inside this framework. The probability schools come into opposition since they adopt philosophical criteria to circumvent the single case problem; instead, the present work applies analytical methods.

**4.4** The next three theorems describe the statuses of physical outcomes during the intervals $T_1$ and $T_2$.
It may be that a scientist discovers that $\mathbf{E}_n$ is an aleatory event from the analysis of its physical characteristics; under this assumption, the *theorem of initial conditions* (TIC) proves that $\omega_n$ is random in $T_1$:

$$\omega_n = \omega_n^{(I)}, \qquad 0 \leq t < t_\omega, \quad n = any. \tag{16}$$

For example, suppose that a rotating urn contains five red balls and five white balls. The mechanical system is not governed by minute rules so the result 'red ball' (or 'white ball') occurs randomly during $n$ rotations no matter the value of $n$.

**4.5** The following theorems detail what happens after $t_\omega$. *The theorem of continuity* (TC) demonstrates that the outcome of the long-term event keeps the indeterminate status in $T_1$ and $T_2$:

$$\omega_\infty = \omega_\infty^{(I)}, \qquad 0 \leq t. \tag{17}$$

For example, suppose a ball is drawn 1,000 times from the urn of the previous case. Both the results turn out to be uncertain in $T_1$; the statistical distributions of the colors show how the results 'red ball' and 'white ball' remain indetermined in $T_2$.

**4.6** *The theorem of discontinuity* (TD) provides the most astonishing result; it proves that the outcome of the single event $\mathbf{E_1}$ switches from the indeterminate to the determinate status at the end of $T_1$:

$$\omega_1^{(I)} \to \omega_1^{(D)}, \qquad t = t_\omega. \tag{18}$$

E.g. A ball is drawn from the rotating urn cited above. The result 'red ball' comes to be determined both when the drawn ball is red and when white. In the first case it becomes certain, in the second impossible.
E.g. Let $\varphi(n)$ the angle between 0 and the generic number $n$ ($0 \leq n \leq 36$). The ball $v$ can occupy a cell in the border of the roulette wheel (Figure 1), thus, the angular position of $v$ verifies:

$$n\left(\frac{360}{37}\right) \leq \varphi(\nu,n) \leq (n+1)\left(\frac{360}{37}\right). \tag{19a}$$

The numbers are equally likely when the wheel is rotating, thus the spatial probability of the ball is:

$$P_1\,\nu = P_1\,\varphi(\nu,n) = P_1\left[n\left(\frac{360}{37}\right) \leq \varphi(\nu,n) \leq (n+1)\left(\frac{360}{37}\right)\right] = 1/37, \quad 0 \leq t < t_n. \tag{19b}$$

When the ball stops, the spatial probability of the extracted number is a unit:

$$P_1\,\nu = P_1\,\varphi(\nu,n) = 1, \quad t_n \leq t. \tag{19c}$$

Eqns. (19b) and (19c) verify the discontinuity (18). This case will become a telling example in quantum physics which uses spatial probability.

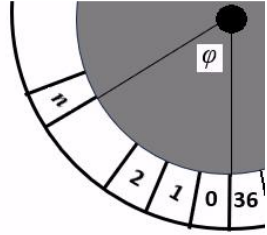

**Figure 1** – Roulette wheel

Probabilists take philosophical positions when facing with the single case problem, instead TSN, TIC and TD provide analytical answers.

**4.7** These theorems show that the probability status $\omega_1^{(I)}$ is real during $T_1$ but cannot be checked, so the single case problem turns out to be a question of testing and not a metaphysical issue.

In consequence of this conclusion, we remark that aside *direct testing methods*, there are *indirect methods* that are very familiar, for example, in astronomy, vulcanology and other experimental fields whose objects cannot be easily accessed. In the probability domain, the impossibility to check $P(E_1)$ can be circumvented if the single event is repeatable. *The corollary of indirect testing* proves that if:

$$\mathbf{E_1} \in \mathbf{E}_\infty. \tag{20}$$

And the long-term event is random and objectively controlled [28][29]:

$$1 < P(\mathbf{E}_\infty) < 0,$$
$$1 < P(\omega_\infty) < 0. \tag{21}$$

Then, also each single trial and result are objectively controlled due to (12) and (13):

$$1 < P(\mathbf{E_1}) < 0,$$
$$1 < P(\omega_1) < 0. \tag{22}$$

This technique allows the practitioner to overcome the single case problem on the empirical level. Note how assumption (20) is not cumbersome because exact sciences (e.g. physics, chemistry, etc.) normally investigate replicable phenomena. In substance, the corollary proves that what is intractable in the individual case can be handled reliably in aggregates of results.

**4.8** Even though we only addressed the single case problem (point **δ**), we can notice that the mathematical approach offers the ensuing advantages over the current literature:

1) The conclusions come from theorems and not from philosophical ruminations or personal decisions.
2) The theorems show how the single case problem regards experimental limitations and can be bypassed using indirect testing methods.
3) The theorems provide analytical descriptions of the single random outcome that becomes determinate and collapses when the event terminates.
4) They cross classical and quantum physics.

**5 Applications in quantum physics**

This work seeks to demonstrate the comprehensiveness of *the structural theory of probability*. For this reason, we apply the definitions and theorems of Sections 3 and 4 to quantum physics, and shall overlook the mathematical formalisms and theories ordinarily employed in QM, such as Hilbert spaces, self-adjoint linear operators, spinors etc. We mean to address: *the wave/particle dualism* (**i**), *the wave collapse* (**ii**) and *the measurement problem* (**iii**) which fragmentary probability theories underpin with difficulty. We neglect the entanglement problem, the Bell's theorem and other questions.

Perhaps the reader doubts the introduction of $P$ into quantum physics because quantum probability $Pr$ has unique properties, e.g. incommutability, negativity etc. The answer is the following.
We will use exclusively the quantities (21) and (22). The integer values and the decimal values have the same meanings for $P$ and $Pr$, therefore $P$ and $Pr$ are consistent in the present inquiry.

**5.1** *Definitions* – Speaking in general, a *computational formula* of $X$ obtained from a mathematical statement, proves to be very useful to calculate the physical parameter or entity $X$, but only the *definitional formula of $X$*, which comes from experience, fixes the intrinsic nature of $X$. E.g. the particle's electric charge $q$ and velocity $v$ give the Faraday force within an electric and magnetic field:

$$F_F = q\ \mathbf{E} + q\ (v \times \mathbf{B}). \tag{23}$$

This equation deriving from the Maxwell's equations along with Lorentz force law, does not expound the general qualities of the force $F$ given by:

$$F = m \cdot a. \tag{24}$$

If one employs (23) in the place of (24) he falls into a web of irresoluble problems, the same occurs when quantum scholars mean to use the wavefunction to explain the double nature of quanta. In fact, $\Psi$ does not derive from experience but from the Schrödinger equation, therefore $\Psi$ is computational and unable to clarify vexed questions. For instance, the Schrödinger equation yields wavefunctions that can be more or less spatially dispersed. The range includes the extreme cases of the Dirac delta (complete localization) and the plane wave (complete delocalization). These results are 'computationally' correct but do not explain what is a wave and a particle from the physical viewpoint. The continuous spectrum does not separate the two material states of the quantum and in a way denies the particle/wave duality. The present research avoids the misuse of $\Psi$ and puts forward two distinct definitions for the particle and the wave based on probability.

The description of a physical entity must descend from experimental observations, and universal experience shows that the quantum $\xi$ is a portion of energy and eventually matter concentrated at one point or otherwise widespread over a certain volume. We assume these topological properties, which are preserved under any quantum transformation, express the essential nature of the particles and waves respectively.
We should employ a spatial-density function but cannot do that since $\xi$ is a discrete quantity. Quanta are indivisible units and we introduce the *spatial probability $P(\xi \in r)$* that is the probability of finding $\xi$ in the point $r = (x,y,z)$ of the Euclidean space $\Sigma$. Using definitions (10) and (11) we posit:

$$\xi \text{ is a particle if it has the determinate status } \xi^{(D)}; \tag{25a}$$
$$\xi \text{ is a wave if it has the indeterminate status } \xi^{(I)}. \tag{25b}$$

From (25a) we infer that the Dirac function $\delta = \delta(r, t)$ with $P(\xi \in r) = 1$ in a point and zero elsewhere, depicts the particle $\xi^{(D)}$ in detail. At the other side, the squared $|\Psi(r, t)|^2$ provides the exact shape of the spatial probability distribution which depends on specific physical constraints. The nature of quantum wave can be compared to the ball of the roulette whose spatial probability distribution is constant along the circumference when the speed of the ball is constant; in parallel, $|\Psi(r, t)|^2$ shows how the energy/matter is dynamically and probabilistically distributed accordingly to the position $r$.

The distributions $|\Psi|^2$ and $\delta$ give the account of energy/mass which are either concentrated or diffused, so they are real for the present theory. Sensors detecting the intensity of incoming energy/mass corroborate the present remark; the possibility of conducting direct or indirect tests will be examined in the next sections.
The statuses $\xi^{(D)}$ and $\xi^{(I)}$ have mutually exclusive probability values and we get:

$$\xi = \xi^{(D)} \text{ OR } \xi^{(I)}. \tag{26}$$

**5.2** *The simplest application* – The *free flight* is the simplest physical phenomenon that is conceptually symmetrical to the linear and constat motion of classical mechanics.
Definition (5) allows us to formalize the free motion this way:

$$\mathbf{E}_\xi = (\alpha_\xi, \rho_\xi, \xi). \tag{27}$$

Where the initial component is the source $\alpha_\xi$ that shots quanta, then the movement $\rho_\xi$ produces the outcome $\xi$. Free flight consists of one or more quanta that maintain their energy/mass nor are they affected by any special effect, such as entanglement, spinning, relativity and others. When something interferes with $\xi$ the flight is no longer 'free' and terminates in a way, namely $\xi$ can continue to move but the free state is no longer there.
A variety of influencing effects can result in the interruption of $\mathbf{E}_\xi$. We confine attention to anelastic collisions between quanta which explain both *microscopic* and *macroscopic* interactions caused by the measurement instruments: sensors, probes etc. The present framework interprets quanta and the measure tools as separate entities in accordance with classical mechanics. Specifically, assuming that the measurement process is destructive, the quantum moves during $T_1$ ($T_1 > 0$) while $T_2$ lasts only an instant.

In order to apply the theorems presented in the past section, we take an ergodic source (e.g. a laser, thermionic tungsten filament, a furnace etc.) as $\alpha_\xi$ which triggers the random flight $\mathbf{E}_\xi$. The theorem of initial conditions concludes that under this conditions $\xi$ is indeterminate, namely $\xi$ is a wave during $T_1$. Thomson's and Davisson–Germer's experiments corroborate this conclusion. In substance, TIC [24] suggests to link the status of the moving quantum to the emitter.

**5.3** In consequence of TLN and TSN, the present theory distinguishes the *single wave* or *wavelet* $\xi_1^{(I)}$ – e.g. laser equipment casting one photon at time – from the *intense wave* or *radiation* $\xi_\infty^{(I)}$ which includes innumerable wavelets due to (13)

$$\xi_\infty^{(I)} = (\xi_{11}^{(I)}, \xi_{12}^{(I)}, \xi_{13}^{(I)}, \ldots). \tag{28}$$

The relation between $\xi_1^{(I)}$ and $\xi_\infty^{(I)}$, here deduced on the theoretical plane, is a normal concern in quantum experiments. For example, operators reduce (or increase) the intensity of the emitter and doing so the intense beam becomes weak (or vice versa). They factually create a single wave at time or a flow of wavelets.

**5.4** Theorems show that when the flight finishes, there are different aftermaths for $\xi_\infty^{(I)}$ and $\xi_1^{(I)}$.

- The theorem of continuity proves that *the radiation $\xi_\infty^{(I)}$ keeps the indeterminate status in $T_2$*. This is amply verified, for example, in classical optics.

- TSN proves that the wavelet $\xi_1^{(I)}$ cannot be directly checked in $T_2$ and the theorem of discontinuity (18) specifies that $\xi_1^{(I)}$ becomes determinate; physically $\xi_1^{(I)}$ becomes *the particle* $\xi_1^{(D)}$. TD prediction is carried out as follows: the collision of the wavelet against the sensor screen (or another measurement detector) causes the free motion to finish, and factually the diffused energy $\xi_1^{(I)}$ condenses in a point. The location of this point over the sensor cannot be forecast since we have assumed $\mathbf{E}_\xi$ is a random process.

The statistical behavior of $K$ wavelets ($K \rightarrow \infty$) approximates $\xi_\infty^{(I)}$ and thus conforms with TLN and TC. The double slits experiment, which we shall see in the next section, offers an example case of these mechanisms.

The structural theory of probability predicts phenomena at both macroscopic and microscopic levels. It adopts uniform concepts and establishes a logical bridge between classical and quantum physics. This is a kind of *classicization* showing how every mechanical event is subject to the same rules.

Numerous tests should verify the current theoretical scheme. Next section discusses the double slit experiment while book [24] analyzes fifteen experiments with quanta moving freely.

**6 The double slit experiment**

Let us look into two versions of the experiment whose ergodic source **A** emits an intense beam and a dim beam, made by several photons and a single photon at time, respectively. In both the versions, photons go through the slits **F** (Figure 2).

► When **A** casts a strong beam of photons, the detector-screen **S** exhibits a continuous pattern in accordance with classical optics.

► When **A** emits a single photon, correspondingly the screen **S** shows one dot. The greater the number of photons sent one by one, the more clearly, they create a discrete pattern on the screen (Figures 3).

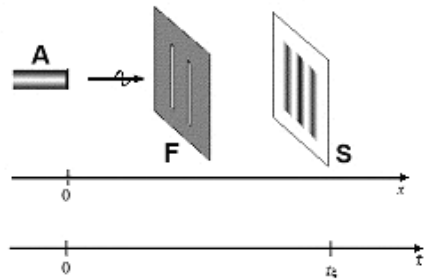

**Figure 2 -** Diagram of the double slit experiment

The intensity of photons detected by **S** is consistent with the concepts of $\xi_\infty^{(D)}$ and $\xi_\infty^{(I)}$ that are real states of energy defined in probabilistic terms. Let us analyze the predictions of the theorems for the two versions of the experiment.

**6.1** *The experiment with the intense beam* – The interference continuous pattern brings evidence that the wave stream $\xi_\infty^{(I)}$ moves in the segment (**A, S**] during the time interval $T_1$ in harmony with TIC:

$$\xi_\infty = \xi_\infty^{(I)}, \qquad 0 \leq t < t_\xi. \tag{29}$$

The pattern also indicates that the indeterminate status $\xi_\infty^{(I)}$ remains in $t_\xi$ – the destructive measurement process causes $T_2$ to last only one instant – and we get:

$$\xi_\infty = \xi_\infty^{(I)}, \qquad t = t_\xi. \tag{30}$$

Equation (30), together with (29), proves that the intense beam keeps the wave state during $T_1$ and $T_2$ and corroborates the theorem of continuity.

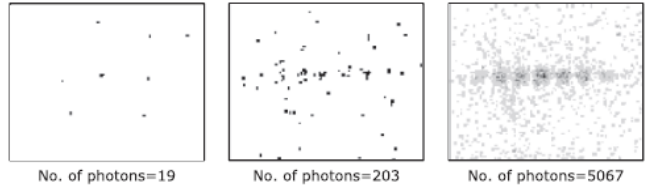

**Figure 3** - Progressive discrete pattern created
by a weak beam of photons through two slits (From [30] figure of public domain)

**6.2** *The experiment with the weak beam* – When **A** emits a photon, the screen exhibits one spot which brings evidence of one particle in $T_2$

$$\xi_1 = \xi_1^{(I)}, \qquad t = t_\xi. \tag{31}$$

When the operator repeats the experiment several times, two effects occur simultaneously.

**6.2.1** The photons emitted one by one cannot interact with one another because of the wide time-space separation interposed between them, and physicists conclude that each incoming photon interferes with itself. That is to say, there is a wavelet in $T_1$:

$$\xi_1 = \xi_1^{(I)}, \qquad 0 \leq t < t_\xi. \tag{32}$$

This evidence corroborates TIC. Joining (31) with (32), we obtain the switching of the single wave that supports the theorem of discontinuity:

$$\xi_1^{(I)} \rightarrow \xi_1^{(D)}, \qquad t = t_\xi. \tag{33}$$

**6.2.2** Expression (28) says that several incoming wavelets make the radiation $\xi_K^{(I)}$ in the long run:

$$\xi_K^{(I)} = (\xi_{11}^{(I)}, \xi_{12}^{(I)}, \xi_{13}^{(I)}, \ldots, \xi_{1K}^{(I)}), \quad K \rightarrow \infty. \tag{34}$$

Empirical data shows that the larger $K$, the more clearly the discrete spectrum comes to sight, and substantiates TLN proving that the greater the number of trials, the closer the empirical intensity approaches the calculations. Moreover, the discrete pattern on **S** brings evidence that the stream $\xi_K^{(I)}$ remains in the wave status in $T_2$, and this detail substantiates the theorem of continuity.

In summary, effect **6.2.1** regards the wavelet that interferes with itself and collapses due to the measurement process. Effect **6.2.2** regards the stream (34) which remains indeterminate during $T_1$ and $T_2$, and creates the discrete spectrum.
No doubt the two versions of the experiment turn out to be rather complex due to various overlapping effects. The following list should aid the reader:
- TIC holds that the incoming beam of photons are waves in **6.1** and **6.2** due to the ergodic source which sets off a non-deterministic movement.
- TLN ensures that an intense flow of photons can be tested, it governs **6.1** and **6.2.2**.
- TSN denies the possibility of the experimental control of a wavelet and regulates **6.2.1**.
- TD predicts the collapse of the wavelet in **6.2.1**.
- TC proves that the waves produced in **6.1** and **6.2.2** remain in $T_2$.

## 7 Discussion and conclusion
This paper includes two parts: Sections 3 and 4 illustrate some traits of the *structural theory of probability* **(A)**; Sections 5 and 6 discuss *the probability-based interpretation of QM* **(B).**

**(A)** The idea that probability confronts a variety of problems, and each theory is not complete guided the preparation of the new construct which begins with the accurate description of the event assessed by $P(\mathbf{E})$. The algebraic structure **E** has been employed to place the frequentist, subjective, logical etc. viewpoints under a unique roof, thus this framework does not center on a particular aspect of indeterminism (**α**); it proves the formulas left to intuition (**γ**) and formalizes the single case problem (**δ**).

Here, the structural theory provides details about the last problem. Specifically, TSN and TD predict the impossibility of direct testing that is an operational obstacle and not a philosophical question.

**(B)** The outcomes presented in **(A)** underpin the following novel answers to the particle/wave duality (**i**), the quantum collapse (**ii**) and measurement problems (**iii**).

   i. Because the wavefunction is not definitional, (25a) and (25b) determine the nature of particles and waves that is under discussion for a long time. TLN and TSN imply the separation of the intense radiation from the single wave.
   ii. TD proves that the wavelet changes status due to the end of the *free motion* and factually the diffused portion of energy/mass condenses in a point of the space.
   iii. The measurement process is the macroscopic action that causes the free motion to finish.

Precise physical features characterize the free movement. As first, it starts with the ergodic source $\alpha_\xi$ that makes $\mathbf{E}_\xi$ to be random and TIC proves how this random movement involves waves. As second, we assume that the random free flight of $\xi_1^{(I)}$ terminates when it bumps and loses energy. Anelastic collisions can be caused by a variety of material entities, so measuring instruments emerge among the most common factors interrupting the free motion.

Schrödinger's equation describes a system which closely 'resembles' the classical, deterministic evolution of a physical system, so quantum scientists ask: Is there a deterministic pre-measurement reality and a post-measurement reality randomly generated by the observer?

The present study shows how the contrary is true. Using an ergodic source, the flight $\mathbf{E}_\xi$ and $\xi$ are random, namely the pre-measurement reality is indeterministic, while the end of $\mathbf{E}_\xi$ causes the output to collapse, namely $\mathbf{E}_\xi$ brings forth a determinate output that is the particle. All this occurs due TD, so the roles of the observer and his consciousness are nonsensical here.

The discrete interference pattern of the double slit experiment reveals that several individual quanta make the flow $\xi_\infty^{(I)}$. Every incoming wavelet breaks down and all together the wavelets approximate the radiation $\xi_\infty^{(I)}$ that can be tested in conformity with TLN. Section 6 illustrates the experimental results obtained with two slits; symmetrical outcomes are obtained with one slit [24]. Bach and colleagues [31] show that when a slit is closed, the weak beam crosses the slit left open and creates a discrete diffraction pattern.

The solutions to problems (**i**), (**ii**) and (**iii**) have the following advantages over current quantum interpretations:

- They descend from theorems valid in both classical and quantum physics, namely, the probability-based interpretation of QM falls within the broad framework of 'the logic of the uncertain' so called by de Finetti.
- They are punctually supported by experiments that give evidence of waves and particles in free motion.
- The explanations conform to intuition and deny the bizarre models circulating in the quantum literature.

The present scheme is limited to the free motion and other phenomena will be investigated later.